\documentclass[11pt]{article}
\usepackage{graphicx}
\title{\bf The inclusive jet production in the BFKL-Bartels approach
with a running coupling introduced via bootstrap}
\author{M.A.Braun\\
Dep. of High Energy physics,
 Saint-Petersburg State University,\\
198504 S.Petersburg, Russia}

\newcommand\lra{\leftrightarrow}
\newcommand\beq{\begin{equation}}
\newcommand\eeq{\end{equation}}
\newcommand\pd{\partial}
\begin{document}

\maketitle

{\bf Abstract}

The inclusive cross-section for production of a jet with a given
transverse momentum off a heavy nucleus is derived in the BFKL
framework with a running coupling on the basis of the bootstrap
relation. The cross-section depends on the same three different
coupling constants as the total cross-section unlike the cross-section
for gluon production derived in the dipole approach.

\section{Introduction}
Attempts to introduce the running coupling into the BFKL dynamics have
a long history. As early as in 1986 L.N.Lipatov  introduced the
running coupling by using an approximate form of the BFKL equation and
a semi-classical approach \cite{lipatov}. In his derivation a single
running coupling $\alpha_s(r)$ appears with $r\to 0$. As a result he
found that the  cut in the complex angular momentum  $J$ transforms
into a sequence of poles, which condensate in a certain region
depending on the total pomeron momentum.
Later in our papers  we introduced the running coupling into the BFKL
equation by imposing the so-called bootstrap condition necessary for the
fulfilment of unitarity ~\cite{braun1,braun2}. In this approach three different
running couplings appear. The structure of the singularities in the
$J$ plane turned out dependent on the assumpsions about the uncontrolled
low energy behaiour of the coupling. Remarkably it turned out that the
bootstrap method correctly reproduced the running of the coupling in
the reggeon interaction both in the forward and non-forward directions
~\cite{vac1,vac2}
explicitly calculated in ~\cite{falip,camcia,fafio}. Still later, with the advent of
the dipole picture and construction of the Balitski-Kovchegov equation for
the pomeron in a heavy nucleus the running coupling was introduced by
explicitly taking into account quark-antiquark loops in the evolution of the
gluon density ~\cite{kovchegov1,weigert,kovchegov2,balitsky}. Again three different coupling constants
appeared in the final equation. Remarkably this procedure turned out
to be fully equivalent to the bootstrap approach, which leads
to the same results with much less labour  ~\cite{braun3}. Finally a few years
ago the dipole approach was generalized to the case of inclusive production
off a heavy nucleus ~\cite{kovchegov3}.
There at the leading order the  number of different
coupling constants proliferated up to seven. Still the authors
made a conjecture for the form of the final inclusive cross-section, in
whih the number of different coupling constants is reduced to three with
one of them depending on the collinearity of the final jet components
$\Lambda _{coll}$, that is essentially on the experimental setup.

In this paper we study the inclusive cross-section off a heavy nucleus
within the bootstrap
approach. It enables us to introduce the running coupling keeping the same
number of them (three) as for the total cross-section.
All the coupling constants remain depending only on the QCD scale $\Lambda$,
with no dependence on the  experimental conditions. Actually this
approach allows to obtain only the inclusive cross-section for the production of a jet
with a given transverse momentum, without specifying its other
characteristics.

 Note that this observable is different from the one calculated in  ~\cite{kovchegov3},
where the jet (or a "particle') not only had   a fixed  transverse
momentum but also had an upper bound on its collinearity $\Lambda _{coll}$.
The latter was assumed to be small
and only the singlular terms terms at $\Lambda _{coll}\to 0$ were kept.
In the case of hadron production it appears that $\Lambda _{coll}$
would become the factorization scale between the hard process and fragmentation function.
In our approach $\Lambda _{coll}$ seems to be effectively taken to infinity, making both the
observable and its calculation very different.

In any case our  cross-section has a well determined physical meaning and can well be
related to experimental observations. So we believe
that it is worth studying, especially since it turns out much simpler than
the conjectured cross-sections in ~\cite{kovchegov3}, which are more specific and
so dependent on the experimental resolution.

In our derivation
we use the well-known two-dimensional transverse picture of BFKL and J.Bartels.
(see e.g.~\cite{BE})
We invoke the relative coefficients between different configurations
in assumed correspondence with the standard AGK
rules, which is confirmed by calculations with
a fixed coupling.

\section{Triple-Pomeron vertex}

As mentioned in the Introduction, in this paper we follow the idea to introduce
a running coupling via the bootstrap ~\cite{braun1,braun2}. Derivation
of the triple-pomeron vertex in the limit $N_c\to\infty$ then goes as
presented in ~\cite{braun4, bravac} for the fixed coupling case.
The same derivation for the running coupling case is presented in
~\cite{braun3}, which we briefly recapitulate here.

Basic formulas for the introduction of a running coupling via the bootstrap
condition consist in expressing both the gluon trajectory $\omega$ and
intergluon interaction in the vacuum channel $V$ in terms of a
single function $\eta(q)$ of the gluon momentum, which then can be chosen
to conform to the high-momentum behaviour of the gluon density with a
running coupling. Explicitly
\beq
\omega(q)=-\frac{1}{2}N_c\int \frac{d^2q_1}{(2\pi)^2}\frac{\eta(q)}
{\eta(q_1)\eta(q_2)},
\label{traj}
\eeq
\beq
V(q_1,q_2|q'_1,q'_2)=N_c\Big[\Big(\frac{\eta(q_1)}{\eta(q'_1)}+
\frac{\eta(q_2)}{\eta(q'_2)}\Big)\frac{1}{\eta(q_1-q'_1)}-
\frac{\eta(q_1+q_2)}{\eta(q'_1)\eta(q'_2)}\Big].
\label{int}
\eeq
In these definitions it is assumed that $q_1+q_2=q'_1+q'_2=q$.
Note that the unsymmetric form (\ref{int}) assumes that the initial
pomeron with momenta $q'_1$ and $q'_2$  is "amputated", that is multiplied
by $\eta(q'_1)\eta(q'_2)$. This factor appears explicitly in the denominators
in (\ref{int}).
For arbitrary $\eta(q)$ the following bootstrap relation is satisfied:
\beq
\frac{1}{2}\int\frac{d^2q'_1}{(2\pi)^2}V(q,q_1,q'_1)=\omega(q)-\omega(q_1)-
\omega(q_2).
\eeq
The fixed coupling corresponds to the choice
\beq
\eta^{fix}(q)=\frac{2\pi}{g^2}q^2.
\label{etafix}
\eeq
Then one finds the standard expression for the trajectory $\omega(q)$
and
\[V^{fix}(q,q_1,q'_1)=\frac{g^2N_c}{2\pi}V_{BFKL}(q,q_1,q'_1),\]
where $V_{BFKL}$ is the standard BFKL interaction. Note that the extra $2\pi$
in the denominator corresponds to the  BFKL weight $1/(2\pi)^3$ in the
momentum integration, which we prefer to take standardly as $1/(2\pi)^2$.

From the high-momentum behaviour of the gluon distribution with a running
coupling one finds
\beq
\eta(q)=\frac{1}{2\pi}bq^2\ln\frac{q^2}{\Lambda^2},\ \ q^2>>\Lambda^2,
\label{asym}
\eeq
where $\Lambda$ is the standard QCD parameter and
\beq
b=\frac{1}{12}\Big(11 N_c-\frac{2}{3}N_f\Big).
\label{bval}
\eeq
As to the behaviour of $\eta(q)$ at small momenta, we shall assume
\beq
\eta(0)=0,
\label{eta0}
\eeq
which guarantees that the gluon trajectory $\omega(q)$
passes through zero at $q=0$ in accordance with the gluon properties.
The asymptotic (\ref{asym}) and condition (\ref{eta0}) are the only
properties of $\eta(q)$ which follow from the theoretical reasoning.
A concrete form of $\eta(q)$ interpolating between (\ref{eta0}) and
(\ref{asym}) may be chosen differently. One hopes that the following
physical results will not strongly depend on the choice.

Our old derivation in ~\cite{braun4} of the triple-pomeron vertex
was actually based on the property (\ref{eta0}) obviously valid for
(\ref{etafix}), the
bootstrap relation and the relation between the Bartels transition
vertex for a fixed coupling constant $W^{fix}_{2\to 3}$ and
intergluon BFKL interaction $V_{BFKL}$  (Eq. (12) in
~\cite{braun4})
\beq
W^{fix}_{2\to 3}(1,2,3|1',3')=V_{BFKL}(2,3|1'-1,3')-V_{BFKL}(12,3|1'3').
\label{oldw}
\eeq
Here and in the following we frequently denote gluon momenta just by numbers:
$1\equiv q_1$, $1'\equiv q'_1$ etc. Also we use $12\equiv q_1+q_2$.
All the rest conclusions were obtained from these three relations in a
purely algebraic manner.

If we {\it define} the transition vertex for a running coupling by a similar
relation in terms of  the new intergluon vertex $V$, Eq. (\ref{int}),
\beq
W_{2\to 3}(1,2,3|1',3')=V(2,3|1'-1,3')-V(12,3|1'3').
\label{new}
\eeq
where $V$ is given by (\ref{int})
then all the results will remain valid also for arbitrary $\eta(q)$
satisfying (\ref{eta0})
and thus for a running coupling, provided $\eta(q)$ is chosen appropriately.
In this way in the momentum space one can find the the three-pomeron vertex with the running
coupling ~\cite{braun3} and write the corresponding Balitski-
Kovchegov equation . It has a simpler form in the cordinate space.
In the forward case at fixed impact parameter $b$
the resulting Balitski-Kovchegov equation with
the running coupling ~\cite{braun3} is obtained as
\[
\frac{\pd}{\pd y}\Phi(y,r)=
\frac{1}{2\pi^2}N_c\int d^2r_2d^2r_3\delta(r-r_1+r_2)
\Big(\frac{\alpha_s(r_1)}{r_1^2}+\frac{\alpha_s(r_2)}{r_2^2}
-2\frac{\alpha_s(r_1)\alpha_s(r_2)}{\alpha_s(r_1,r_2)}
\frac{{\bf r}_1{\bf r}_2}{r_1^2r_2^2}\Big)\]\beq
\Big(\Phi(y,r_1,b)+\Phi(y,r_2,b)-\Phi(y,r,b)-
\Phi(y,r_1,b)\Phi(y,r_2,b)\Big).
\label{forintr2}
\eeq
where the two running coupling constants $\alpha_s(r)$ and
$\alpha_s(r_1,r_2)$ are defined as
\beq
\alpha_s(r)=-\pi^2 r^2 f(r,r),\ \ \alpha_s(r_1,r_2)=
-\pi^2({\bf r}_1{\bf r}_2)\frac{f(r_1,r_1)f(r_2,r_2)}{f(r_1,r_2)}
\eeq
and
\beq
f(r_1,r_2)=\int d^2\rho
\tilde{\eta}(\rho)\xi(r_1-\rho)\xi(r_2-\rho).
\label{ffun1}
\eeq
where $\tilde{\eta}(r)$ is the Fourier transform of $\eta(q)$ and
$\xi(r)$ is the Fourier transform of $1/\eta(q)$.
Eq. (\ref{forintr2}) formally coincides with the running coupling equation
obtained in the dipole aproach in ~\cite{kovchegov1,kovchegov2}.
However in our approach the couplings $\alpha_s(r)$ and $\alpha_s(r_1,r_2)$
are not directly taken from the QCD. Rather it is function $\eta(q)$ that
is taken from the QCD and the coupling constants are determined by it.

\section{Inclusive cross-sections}
Diagrammatically contributions to the forward amplitude for the
collision of the projectile with the nucleus are identical to those
which appear in the fixed coupling case. The difference is totally
restricted to the new form of the reggeized gluon trajectory,
interaction between the reggeons and the splitting vertex.
Correspondingly the inclusive cross-sections will be obtained either
by cutting the interaction in the uppermost pomeron (before all splitting)
or by cutting the first splitting vertex. All the rest contributions
will be eliminated by the AGK cancellations.

\begin{figure}
\begin{center}
\includegraphics[scale=0.65]{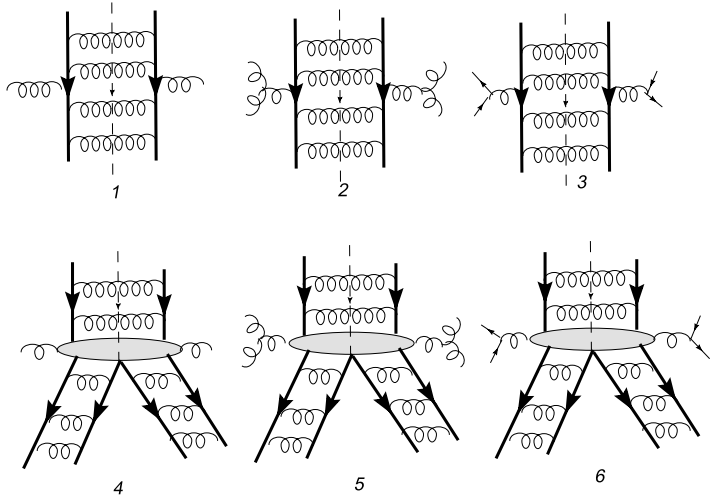}
\end{center}
\caption{Diagrams for the cross-section corresponding to emission from the upper pomeron (1-3)
and from the triple pomeron vertex (4-6). Thick solid lines denote reggeons, thin lines
denote real quarks, wavy lines denote real and virtual gluons. }
\label{fig1a}
\end{figure}

However one should take into account that the inclusive cross-sections
obtained in this manner do not refer to precisely gluon production.
The form of the function $\eta(p)$ desribing the intermediate $s$-channel
state with the single-loop $\beta$ function includued implies that one
has to take into account not only the single real gluon state but also
states which contribute to this $\beta$-function, namely the quark-antiquark
and two-gluon states. So the inclusive cross-section obtained by
fixing the intermediate $s$-state momentum $p$ actually refers to all possible
states having this momentum. It does not discriminate between contributions
from gluons (single or in pairs) and (anti)quarks. So it is rather the
inclusive cross-section for emission of a jet with a total transverse
momentum $p$.  The diagrams for this cross-section are illustrated
in Fig. \ref{fig1a}.
We think that this quantity has a well-defined physical
meaning and is accessible for experimental observation and so worth theoretical
investigation. In all the following we study precisiely this generalized
inclusive cross-section for jet production.

\section{Jet production from the pomeron}
This part of the total cross-section is calculated in full analogy
with the fixed coupling case. If only a single pomeron is exchanged
between the projectile and target  then  we have the impulse
approximation contribution
\beq
I(Y,y,p)=\frac{(2\pi)^3d\sigma}{dyd^2p}=\frac{4N_c}{\eta(p)}
\int \frac{d^2k}{(2\pi)^2}\eta(k)\eta(p-k)
P(Y-y,p-k)P(y,k).
\label{ipom}
\eeq
Here it is assumed that the "semi-amputated" forward pomeron wave
function $\phi(y,p)=\eta(p)P(y,p)$
in the momentum space satisfies the equation
\beq
\frac{\partial \phi(y,p)}{\partial y}=2\omega(p)\phi(y,p)+
2N_c\int \frac{d^2k}{(2\pi)^2}\frac{P(y,k)}{\eta(p-k)}\phi(y,k).
\eeq
With a fixed coupling $P(y,p)$ obviously has order $g^4$ so that
$I$ has order $g^6$, which takes into account  two impact factors each
of order $g^2$

For a nucleus target the pomeron coupled to the target has to be substituted
by the solution of the Balitski-Kovchegov equation (\ref{forintr2})
at fixed impact parameter $b$ (transformed to the momentum space):
\beq
I_1(Y,y,p,b)=\frac{(2\pi)^3d\sigma_1}{dyd^2pd^2b}=
\frac{4N_c}{\eta(p)}\int \frac{d^2k}{(2\pi)^2}\eta(k)\eta(p-k)
P(Y-y,p-k)\Phi(y,k,b).
\label{i1}
\eeq

\section{Jet production from the vertex: generalities}
The three-pomeron vertex has a fixed rapidity and does not include
evolution. This makes it possible to study the contribution to
jet emission from the vertex in the lowest order of perturbation theory,
that is for the target consisting of only two centers.
It also allows to simplify treatment choosing for the projectile and
target quarks, modeling the pomeron exchanges by colorless double
reggeon exchanges.
The inclusive cross-section will be obtained from the diagrams for
the forward scattering off two centers, in which reggeon interactions
and splittings are described by functions $V$ and $W$, given by
Eqs. (\ref{int}) and (\ref{new}) with the running coupling.
All the diagrams may be divided into two configurations depending
on the way the four final reggeons are combined into pomerons.
The diffractive one
includes diagrams with two consecutive colourless exchanges (Figs.
\ref{fig1}-\ref{fig3}). The non-diffractive configuration includes
diagrams with parallel colourless exchanges (Figs. \ref{fig4}-\ref{fig6})
when  one of the colourless pair of reggeons is enclosed in the other.
In these diagrams vertical lines denote reggeons, solid horizontal lines
denote projectile and target quarks and wavy lines denote $s$-channel
gluons.
Diagrams in which the colourless pairs partially overlap do not give contribution
to the inclusive cross-section and we do not show them.
Note that the number of reggeons coupled to the projectile may vary from
two to four. However, as shown long ago, all contributions
reduce to that of a colourless pair of reggeons forming the initial pomeron
(amputated with the forms (\ref{int}) and (\ref{new})).
If one subtracts the contribution from the so-called reggeized term
(see e.g. ~\cite{BE}) or alternatively from the Glauber initial condition
in the dipole formalism) the the rest gives the contribution from the
triple pomeron vertex, which is our goal

To obtain the inclusive cross-section for the production of a jet with
momentum $p$ one has to fix function $\eta(p)$ in  intermediate
states in the $s$ channel. These intermediate states are obtained by
cutting the diagrams in the $s$-channel. Different cuts may pass through
one of the targets (single cuts, S), through both targets (double
cuts, DC) or do not pass through targets at all (diffractive cuts, D).
In the diffractive configuration only diffractive and single cuts are
possible. In the non-diffractive configuration only single and double cuts
are possible.
According to the AGK rules the relative weights of the contributions
from diffractive, single and double cuts are $1:-1:2$.

We denote the reggeon momenta of the first final pomeron as $q_1$ and $q_2$
and of the second as $q_3$ and $q_4$. In the forward direction $q_2=-q_1$
and $q_3=-q_4$. So the final pomerons carry momenta $q_1$ and $q_4$.

\section{A. Diffractive configuration}
We divide all contributions according to the number of initial reggeons (R):
A2, A3 and A4 for two three and four initial reggeons respectively.
\subsection{A4: 4R$\to$ 4R}
\begin{figure}
\begin{center}
\includegraphics[scale=0.65]{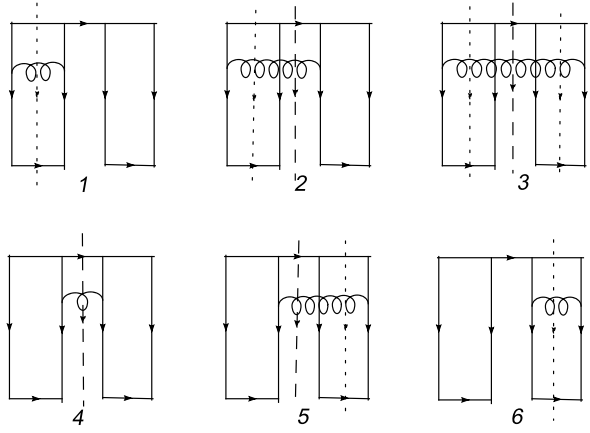}
\end{center}
\caption{4R$\to$ 4R in the diffractive configuration}
\label{fig1}
\end{figure}

We have 6 diagrams shown in Fig. \ref{fig1}.
They contribute either in D or S configurations. Due to assumed relation
between D and S diagrams 2 and 5 are ancelled. 1 and 6 give only S.
4 contains only D.
Finally 3 contains D+2S=-D. Due to the properies of the pomeron
$SD_1=SD_6=0$. So we are left only with 3 and 4.
Colour factors:
\[
C_3=C_4=\frac{1}{4}N_c^4.
\]
So taking $(-V)$ for the interaction we find
\beq
(3)=-C_3(-V(q_1,q_4|q_1+p,q_4-p)),
\eeq
\beq
(4)=C_4(-V(-q_1,-q_4|-q_1+p,-q_4+p)).
\eeq
Changing signs of $q_1$ and $q_4$ does not change $V$.
So in the end (3)+(4)=0 and the diagrams $D_{4\to4}$ give no contribution.

\subsection{A2: 2R$\to$ 4R}
\begin{figure}
\begin{center}
\includegraphics[scale=0.65]{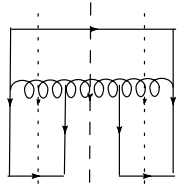}
\end{center}
\caption{2R$\to$ 4R in the diffractive configuration}
\label{fig2}
\end{figure}

There is a single diagram 2R$\to$4R, shown in Fig. \ref{fig2}.
It contributes in both D and S configurations.
Colour factor is $C=N_c^4$.
The D contribution is
\beq
A_{2D}=-CW(q_1,-q_1-q_4,q_4|p,-p).
\eeq
The two S contributions give
\beq
A_{2SD}=+CW(q_1,-q_1-q_4,q_4|q_1+p,-q_1-p)+
CW(q_1,-q_1-q_4,q_4|-q_4+p,q_4-p).
\eeq

\subsection{A3: 3R$\to$ 4R}
\begin{figure}
\begin{center}
\includegraphics[scale=0.65]{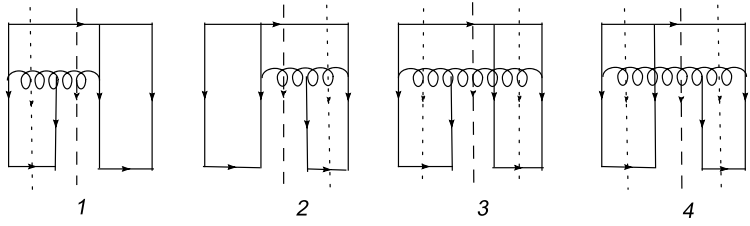}
\end{center}
\caption{3R$\to$4R in the diffractive configuration}
\label{fig3}
\end{figure}

We have 4 diagrams which contribute in $D$ or $SD$ configurations,
shown in Fig. \ref{fig3}
Note that $C_1=C_2=-1/2$ and $C_3=C_4=+1/2$.

So the S contribution from (1) cancels the left
S contribution from (3) and the S conttibution from 2 cancels the
right contribution from (4).  Moreover the D contribution from (3) cancels
the right SD contribution from (3) and the D contribution from (4)
canceles the left S contribution drom (4).
As a result the only remaining contribution is  D contribution
from diagrams (1) and (2)
\beq
A3=A3_D=-C_1W(q_1,-q_1,-q_4|p,-q_4-p)-C_1W(-q_1,-q_4,q_4|-q_1+p,-p),
\eeq
where we have taken into account that D contributions from
(1) and (2) are equal.

\subsection{The total diffractive contribution}
Suppressing the overall colour coefficient $N_c^4$ we have found
\[
A=A_{2D}+A_{2S}+A_{3D}=
-W(q_1,-q_1-q_4,q_4|p,-p)\]\[+W(q_1,-q_1-q_4,q_4|q_1+p,-q_1-p)
+W(q_1,-q_1-q_4,q_4|-q_4+p,q_4-p)
\]
\[
-\frac{1}{2}W(q_1,-q_1,-q_4|p,-q_4-p)
-\frac{1}{2}W(-q_1,-q_4,q_4|-q_1+p,-p).
\]

\section{B. Non-diffractive configuration}
\subsection{B4: 4R$\to$ 4R}
\begin{figure}
\begin{center}
\includegraphics[scale=0.65]{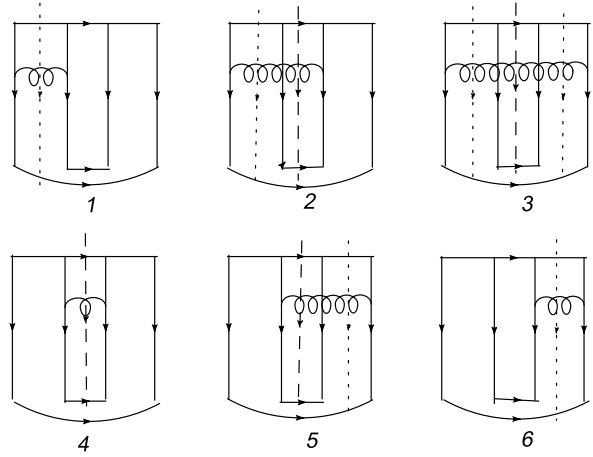}
\end{center}
\caption{4R$\to$ 4R in the non-diffractive configuration}
\label{fig4}
\end{figure}

Six diagrams with transitions 4R$\to$ 4R are shown in Fig. \ref{fig4}.
Since the double cut (DC) contribuition enters with coefficient 2
diagram (3) is cancelled.
Diagrams 2 and 5 give the same contribution. The S in both cancels one
half of the DC contribution, so that the total coefficient is reduced to 1.
Diagram (4) is zero. Diagram 1 and 6 contribute only to the S contribution.
So in the end the DC and S contributions are
\beq
B4_{DC}=(2)+(5),\ \ B4_{S}=-(1)-(6).
\eeq
Colour factors are
\[ C_2=C_5=-\frac{1}{4}N_c^4,\ \ C_1=c_6=\frac{1}{4}N_c^4.\]
So we get (again suppressing $N_c^4$)
\beq
B4_{DC}=\frac{1}{4}\Big(V(-q_4,-q_1|-q_4+p,-q_1-p),\ \
+V(q_1,q_4|q_1+p,q_4-p)\Big),
\eeq
\beq
B4_{S}=\frac{1}{4}\Big(V(-q_4,q_1|-q_4+p,q_1-p)
+V(-q_1,q_4|-q_1+p,q_4-p)\Big).
\eeq

\subsection{B2: 2R$\to$ 4R}
\begin{figure}
\begin{center}
\includegraphics[scale=0.65]{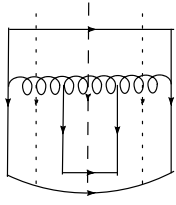}
\end{center}
\caption{2R$\to$ 4R in the non-diffractive configuration}
\label{fig5}
\end{figure}

There is a single diagram for transitions 2$\to 4$ shown in Fig. \ref{fig5}.
It contains DC and S contributions. The colour factor is $C=N^4$,
so suppressing it
\beq
B2_{DC}=-2W(-q_4,0,q_4|q_1-q_4+p,-q_1+q_4-p)
\eeq
and
\beq
B2_{S}=2W(-q_4,0,q_4|-q_4+p,q_4-p).
\eeq
In fact the S contribution gives zero: it does not depend on $q_1$.

\subsection{B3: 3R$\to$ 4R}
\begin{figure}
\begin{center}
\includegraphics[scale=0.65]{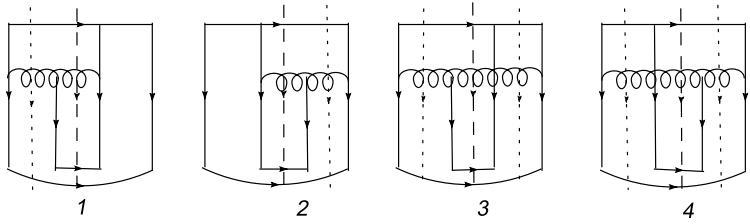}
\end{center}
\caption{3R$\to$ 4R in the non-diffractive configuration}
\label{fig6}
\end{figure}

Four diagrams with transitions 3$\to$4 are shown in Fig. \ref{fig6}.
We observe that the right S contribution in (3) cancels one half of the DC
contribution in (3) and the left S contribution in (4) cancels one
half of the DC contribution in (3).
Thus we get
\beq
B3_{DC}=2(1)_{DC}+2(2)_{DC}+(3)_{DC}+(4)_{DC}
\eeq
and
\beq
B3_{S}=-(1)_{S}-(2)_{S}-(3)_{Sl}-(4)_{Sr},
\eeq
where $Sl$ and $Sr$ refer to left and right single cuts.

The colour factors are
\[C_1=C_2=\frac{1}{2}N_c^4,\ \ C_3=C_4=-\frac{1}{4}N_c^4.\]
So suppressing $N_c^4$
\[
B3_{DC}=-W(-q_4,q_1,-q_1|-q_4+q_1+p,-q_1-p)+
-W(q_1,-q_1,q_4|q_1+p,q_4-q_1-p)\]\[+
+\frac{1}{4}W(-q_4,q_1,q_4|q_1-q_4+p,q_4-p)
+\frac{1}{4}W(-q_4,-q_1,q_4|-q_4+p,q_4-q_1-p)
\]
and
\[
B3_{S}=\frac{1}{2}W(-q_4,q_1,-q_1|-q_4+p,-p)
\frac{1}{2}W(q_1,-q_1,q_4|p,q_4-p)\]\[
-\frac{1}{4}W(-q_4,q_1,q_4|-q_4+p,q_1+q_4-p)
-\frac{1}{4}W(-q_4,-q_1,q_4|-q_1-q_4+p,q_4-p).
\]

\subsection{The total non-diffractive contribution}
\[
B=B4_{DC}+B4_{SDC}+B2_{DC}+B2_{S}+B3_{DC}+B3_{S}=\]
\[
\frac{1}{4}\Big(V(-q_4,-q_1|-q_4+p,-q_1-p)
+V(q_1,q_4|q_1+p,q_4-p)\]\[
+(V(-q_4,q_1|-q_4+p,q_1-p)
+V(-q_1,q_4|-q_1+p,q_4-p)\Big)\]
\[
-2W(-q_4,0,q_4|q_1-q_4+p,-q_1+q_4-p)\]
\[
-W(-q_4,q_1,-q_1|-q_4+q_1+p,-q_1-p)-
W(q_1,-q_1,q_4|q_1+p,q_4-q_1-p)\]\[+
\frac{1}{4}W(-q_4,q_1,q_4|q_1-q_4+p,q_4-p)
+\frac{1}{4}W(-q_4,-q_1,q_4|-q_4+p,q_4-q_1-p)\]
\[
+\frac{1}{2}W(-q_4,q_1,-q_1|-q_4+p,-p)
+\frac{1}{2}W(q_1,-q_1,q_4|p,q_4-p)\]\[
-\frac{1}{4}W(-q_4,q_1,q_4|-q_4+p,q_1+q_4-p)
-\frac{1}{4}W(-q_4,-q_1,q_4|-q_1-q_4+p,q_4-p).
\]

\section{Jet production from the vertex: the final form}
The found contributions to inclusive cross-sections from the two
configurations are numerous and contain many terms.
To analyze them it is convenient to introduce function
\beq
F(q|p)\equiv(q|p)=\frac{\eta(q)}{\eta(p)\eta(q-p)}=(q|q-p)=(-q|-p).
\label{deff}
\eeq
In terms of this function
\beq
V(q_1,q_2|k_1,k_2)=(k_{12}|k_1)-(q_1|k_1)-(q_2|k_2),
\eeq
\beq
W(q_1,q_2,q_3|k_1,k_2)=(k_{12}|k_1)-(q_{12}|k_1)-(q_{23}|k_2)+(q_2|k_1-q_1),
\eeq
where $k_{12}=k_1+k_2$ etc.
The analysis of the found inclusive cross-section in terms of function $F(q|p)$
is performed in Appendix. It is important to take into account that all
contributions which do not depend on the momenta of one of the final
pomerons, $q_1$ or $q_4$, vanish, since the Pomeron vanishes when its
two reggeized gluons are at the same spatial point.

After some cancellations one finally finds the contribution
\beq
A+B=2\Big[(q_{14}|p)+(q_{14}|-p)\Big].
\label{ab}
\eeq
With the fixed coupling constant one finds instead
\beq
(A+B)^{fix}=\frac{g^2}{\pi}
\Big[\frac{q_{14}^2}{p^2(q_{14}-p)^2}+\Big(p\to -p\Big)\Big].
\label{abfix}
\eeq
It was shown that in the fixed coupling case half of this
expression comes from the reggeized piece (or the Glauber initial
condition in the dipole approach) ~\cite{braun5}. By the same reasoning the
contribution
from the triple pomeron vertex is one half of (\ref{ab}).
As we see, to find  our inclusive cross-section for jet production
with the running couling all the necessary change is the substitution
of all momenta according to the rule
\beq
q^2\to\frac{g^2}{2\pi}\eta(q).
\label{rule}
\eeq
Using the known form of the inclusive cross-section in the
fixed coupling case (\cite{KT}, see also \cite{braun5} for closer
notations) one can immediately
write the final expression for it with the running coupling as
\beq
I_2(Y,y,p)=-\frac{2N_c}{\eta(p)}\int\frac{d^2q_1d_2q_4}{(2\pi)^4}
\eta(q_{14}-p)P(Y-y,q_{14}-p)\eta(q_{14})\Phi(y,q_1)\Phi(y,q_4).
\label{i2}
\eeq
The total inclusive cross-section is the sum of (\ref{i1}) and (\ref{i2}):
\[
I(Y,y,p,b)=\frac{(2\pi)^3d\sigma}{dyd^2pd^2b}=
\frac{2N_c}{\eta(p)}\int \frac{d^2q_1d^2q_4}{(2\pi)^4}\]\beq\eta(q_(14-p)
P(Y-y,q_{14}-p)\Big[2\eta(q_1)\Phi(y,q_1,b))(2\pi)^2\delta(q_4)-
\eta(q_{14})\Phi(y,q_1)\Phi(y,q_4)\Big].
\label{itot}
\eeq

\section{Discussion}
Our final expression for the inclusive cross-section Eq. (\ref{itot})
has a strong similarity with the one conjectured in ~\cite{kovchegov3}).
It contains three factors $\eta$, which can be put in correspondence
with the three coupling constants depending on different arguments
in ~\cite{kovchegov3}. However in our formula the arguments of functions
$\eta$ directly depend on the three momenta involved: that of the
observed jet and two of the gluon distributions involved. In contrast,
in the conjecture of ~\cite{kovchegov3} the argument of one of the coupling constants
depends only on the assumed value of collinearity of the observed jet and
the arguments of two others are complex and
depend on all three momenta in a very complicated manner. However,
as mentioned in  the Introduction, the
literal comparison of the two cross-sections is not possible, since in fact
they refer to different processes: to jet production with a given momentum
in our case and with additional restriction on jet collinearity in
~\cite{kovchegov3}. Besides the cross-section in ~\cite{kovchegov3} is
after all only conjecture, whereas ours is more or less consitently derived
from the bootstrap condition which demonstrated its validity for the total
cross-section.

It remains to be seen by practical applications to what extent
this difference is felt in the actual inclusive cross-sections.
To do this consistently one has to previously solve our equation for the
unintegrated gluon density (\ref{forintr2}). With all its similarity to
the currently used equation in the dipole picture, the actual values
of the three running couplings involved are not identical, so that the
already found solutions in the dipole picture cannot be directly used for
our inclusive cross-section.
We postpone this problem for future studies.

In conclusion we stress that
that our  equations take in account terms of the
order $(\alpha(q))^n$ with $\alpha(q)$ taken in the leading order.
Subleading terms of the relative order $1/\ln(q^2/\Lambda^2)$ remain
undetermined, since they correspond to the next-to-leading order in the
running coupling.

\section{Appendix.Analysis of the found inclusive cross-sections}

In terms of $F$ we find
\[
A=
-\Big(0-(-q_4|p)-(-q_1|-p)+(-q_1-q_4|p-q_1)\Big)\]\[
+\Big(0-(-q_4|q_1+p)-(-q_1|-q_1-p)+(-q_1-q_4|p)\Big)\]\[
+\Big(0-(-q_4|-q_4+p)-(-q_1|q_4-p)+(-q_1-q_4|-p)\Big)\]
\[
\frac{1}{2}\Big((-q_4|p)-(-q_1-q_4|-q_4-p)+(-q_1|p-q_1)\Big)
\frac{1}{2}\Big((-q_1|-p)-(-q_1-q_4|-q_1+p)+(-q_4|p)\Big)
\]
We  take into account that terms which do not depend on $q_1$ or do not
depend on $q_4$ are integrated ober $q_1$ and $q_4$ to give zero.
Dropping these terms
\[
A=
-(-q_1-q_4|p-q_1)
-(-q_4|q_1+p)+(-q_1-q_4|p)\]\[
-(-q_1|q_4-p)+(-q_1-q_4|-p)
\]
\[
-\frac{1}{2}(-q_1-q_4|-q_4-p)
-\frac{1}{2}(-q_1-q_4|-q_1+p)
\]

Now non-diffractive contributions
\[B=
\frac{1}{4}\Big((-q_4-q_1|-q_4+p)-(-q_4|-q_4+p)-(-q_1|-q_1-p)\]\[
+(q_1+q_4|q_1+p)-(q_1|q_1+p)-(q_4|q_4-p)\]\[
+(q_1-q_4|-q_4+p)-(-q_4|-q_4+p)-(q_1|q_1-p)\]\[
+(q_4-q_1|-q_1+p)(-q_1|-q_1+p)-(q_4|q_4-p)\Big)\]
\[
-2\Big(-(-q_4|q_1-q_4+p)-(q_4|-q_1+q_4-p)\Big)\]
\[
-\Big((-q_4|-q_1-p)-(q_1-q_4|q_1-q_4+p)+(q_1|q_1+p)\Big)\]\[
-\Big((q_4|q_1+p)-(q_4-q_1|q_4-q_1-p)+(-q_1|p)\Big)\]\[
+\frac{1}{4}\Big((q_1|q_4-p)-(q_1-q_4|q_1-q_4+p)-(q_1+q_4|q_4-p)
+(q_1|q_1+p)\Big)\]\[
+\frac{1}{4}\Big((-q_1|-q_4+p)-(-q_1-q_4|-q_4+p)-(q_4-q_1|q_4-q_1-p)
+(-q_1|p)\Big)\]
\[
+\frac{1}{2}\Big((-q_4|-p)-(q_1-q_4|-q_4+p)+(q_1|-p)\Big)
+\frac{1}{2}\Big((q_4|p)-(q_4-q_1|q_4-p)+(-q_1|-p)\Big)\]\[
-\frac{1}{4}\Big((q_1|-q_4+p)-(q_1-q_4|-q_4+p)-(q_1+q_4|q_1+q_4-p)
(q_1|p)\Big)\]\[
-\frac{1}{4}\Big((-q_1|q_4-p)-(-q_1-q_4|-q_1-q_4+p)-(q_4-q_1|q_4-p)
+(-q_1|-p)\Big)
\]

Again we remove terms which do not depend on one of the variables $q_1$
or $q_4$. Dropping these terms
\[B=
\frac{1}{4}\Big((-q_4-q_1|-q_4+p)+(q_1+q_4|q_1+p)
+(q_1-q_4|-q_4+p)+(q_4-q_1|-q_1+p)\Big)\]
\[
-2\Big(-(-q_4|q_1-q_4+p)-(q_4|-q_1+q_4-p)\Big)\]
\[
-\Big((-q_4|-q_1-p)-(q_1-q_4|q_1-q_4+p)\Big)
-\Big(-(q_4-q_1|q_4-q_1-p)+(q_4|q_1+p)\Big)\]\[
+\frac{1}{4}\Big((q_1|q_4-p)-(q_1-q_4|q_1-q_4+p)-(q_1+q_4|q_4-p)
\Big)\]\[
+\frac{1}{4}\Big((-q_1|-q_4+p)-(-q_1-q_4|-q_4+p)-(q_4-q_1|q_4-q_1-p)
\Big)\]
\[
+\frac{1}{2}\Big(-(q_1-q_4|-q_4+p)\Big)
+\frac{1}{2}\Big(-(q_4-q_1|q_4-p)\Big)\]\[
-\frac{1}{4}\Big((q_1|-q_4+p)-(q_1-q_4|-q_4+p)-(q_1+q_4|q_1+q_4-p)
\Big)\]\[
-\frac{1}{4}\Big((-q_1|q_4-p)-(-q_1-q_4|-q_1-q_4+p)-(q_4-q_1|q_4-p)
\Big)
\]

Next we simplify our expressions using the possibility to change $q_{1,4}
\lra -q_{1,4}$ and $q_{1,4}\lra q_{4,1}$.
We reduce all terms to two basic structures $(q_{14}|a)$ and $(q_1|a)$.
Then we get
\[
A=
-(q_{14}|q_1+p)
-(q_1|q_4+p)+(q_{14}|p)
-(q_1|q_4-p)+(q_{14}|-p)
\]\[
-\frac{1}{2}(q_{14}|q_1+p)
-\frac{1}{2}(q_{14}|q_1+p)
\]
and
\[B=
(q_{14}|q_1+p)
-2\Big(-(q_1|q_4-p)-(q_1|q_4+p)\Big)\]
\[
-(q_1|q_4-p)+(q_{14}|-p)+(q_{14}|p)-(q_1|q_4+p)\]\[
+\frac{1}{4}\Big((q_1|q_4-p)-(q_{14}|-p)-(q_{14}|q_1+p)
\Big)\]\[
+\frac{1}{4}\Big((q_1|q_4+p)-(q_{14}|q_1+p)-(q_{14}|p)
\Big)
-(q_{14}|q_1+p)\]\[
-\frac{1}{4}\Big((q_1|q_4+p)-(q_{14}|q_1+p)-(q_{14}|p)\Big)\]
\[
-\frac{1}{4}\Big((q_1|q_4-p)-(q_{14}|-p)-(q_{14}|q_1+p)
\Big)
\]

Note that
\[(q_{14}|q_1-p)=(q_{14}|q_4-p)=(q_{14}|q_1+p)\]

Let us find the coefficients for our basic structures
\[(q_{14}|q_1+p)\]
A2:-1, A3:+1, B4:+1,B2: 0,B3:-1/4-1/4-1+1/4+1/4=-1,
total=0.\\
\[(q_1|q_4-p)\]
A2:-1, A3:0, B4:0, B2:2, B3:-1+1/4-1/4=-1, total=0.
\[(q_1|q_4+p)\]
A2:-1, A3:0, B4:0, B2:+2, B3:-1+1/4-1/4=-1, total=0.
\[(q_{14}|p)\]
A2:+1, A3:0, B4:0, B2:0, B3:+1-1/4+1/4=+1, total =2.
\[(q_{14}|-p)\]
A2:+1, A3:0, B4:0, B2:0, B3:+1-1/4+1/4=1, total =2.

This brings us to the final result (\ref{ab}).

\section{Acknowledgements}
The author would like to thank Yu. Kovchegov for numerous informative
discussions. This work has been supported by grants RFFI 11.15.360.2015
and SPSU 11,38.223.2015.


\begin{thebibliography}{99}
%
\bibitem{lipatov}Zh.Eksp.Teor.Fiz {\bf 90} (1986) 1536
(Sov.Phys. JETP {\bf 63} (1986) 904).
%
\bibitem{braun1} M.A.Braun, Phys.Lett. {\bf B 345} (1995) 155.
%
\bibitem{braun2} M.A.Braun, Phys.Lett. {\bf B 348} (1995) 190.
%
\bibitem{vac1} M.A.Braun, G.P.Vacca, Phys. Lett. {\bf  B 454} (1999)
319
%
\bibitem{vac2} M.A.Braun, G.P.Vacca, Phys. Lett. {\bf  B 477} (2000)
156
%
\bibitem{falip} V.S.Fadin, L.N.Lipatov, Phys. Lett. {\bf B 429}(1998)
127
%
\bibitem{camcia} G.Camici, M.Ciafaloni, Phys. Lett. {\bf B 395} (1997)
118.
%
\bibitem{fafio} V.S.Fadin, R.Fiore, A.Papa, Phys. Rev. {\bf D 60} (1999)
074025.
%
\bibitem{kovchegov1} Yu.V.Kovchegov and H.Weigert, hep-ph/0609090.
%
\bibitem{weigert} E.Gardi, J.Kuokkanen, K.Rummukainen, H.Weigert,
Nucl. Phys. {\bf A 784} (2007) 282
%
\bibitem{kovchegov2}  Yu.V.Kovchegov and H.Weigert, hep-ph/0612071.
%
\bibitem{balitsky} I.I.Balitsky, Phys. Rev. {\bf D 75} (2007) 014001
%
\bibitem{braun3} M.A.Braun, Eur. Phys. J. {\bf C 51} (2007) 625.
%
\bibitem{kovchegov3} W.A.Horowitz and Yu.V.Kovchegov, Nucl. Phys.
{\bf A 849} (2011) 72.
%
\bibitem{BE} J.Bartels and C.Ewerz, JHEP {\bf 9909} (1999) 026
%
\bibitem{braun4} M.A.Braun, Eur. Phys. J {\bf C 6} (1999) 321.
%
\bibitem{bravac} M.A.Braun and G.P.Vacca, Eur. Phys. J. {\bf C 6} (1999) 147.
%
\bibitem{KT} Yu.V.Kovchegov and K.Tuchin, Phys. Rev. {\bf D 65} (2002)
074026.
%
\bibitem{braun5} M.A.Braun, Eur. Phys. J {\bf C 48} (2006) 501.
%
%
(1999) 319; {\bf B 447} (2000) 156.
%
074025.
%
%
%
(1999) 114036.
%

\end{thebibliography}
\end{document}